\documentclass[preprint,review,12pt]{elsarticle}

\usepackage[hidelinks]{hyperref}
\usepackage{graphicx}
\usepackage{tikz}
\usetikzlibrary{positioning,shapes.geometric,arrows.meta,fit,backgrounds,calc}
\usepackage{booktabs}
\usepackage{longtable}
\usepackage{float}
\usepackage{lmodern}
\usepackage[T1]{fontenc}
\usepackage[utf8]{inputenc}
\begin{document}

\begin{frontmatter}

\title{Relic and Gumi: Governing Memory, Safety, and Continuity in Diegetic Relational Agents}

\author{Daniele Ver\`i\fnref{auth}}
\affiliation{organization={Independent Researcher}, city={Padova}, country={Italy}}
\fntext[auth]{Email: daniele.veri.pe@gmail.com. ORCID: \url{https://orcid.org/0009-0005-1056-8411}.}

\begin{abstract}
Longitudinal relational agents depend on memory. The same memory that makes an agent feel continuous can also make the agent intrusive, clinically overconfident, emotionally manipulative, or opaque to researchers and participants. This paper presents Relic, a governance architecture for diegetic relational agents, and Gumi, a subject-specific companion agent initialized and governed through Relic. The system is designed around a separation between what the system can observe, what researchers can review, what Gumi can remember, and what Gumi can say. We describe three core mechanisms. First, sensitive safety signals are represented as researcher-facing governance objects rather than user-facing explanations or agent memories. Second, Shared Continuity Memory is designed around relational markers grounded in subject-confirmed words, corrections, follow-up permissions, and recall limits. Third, we specify a runtime mediation layer for delivery, proactivity, session identity, output transformation, and platform allowlists; adoption across live paths remains partial. Together, these mechanisms are intended to support relational continuity while reducing the risk that Gumi is configured as a clinical assistant, a mood tracker, a generic assistant, or an opaque profiling system. We contribute a system architecture, object model, researcher workbench design, and proposed evaluation framework for studying governed relational memory in diegetic agents. We also report a small exploratory pilot with the author and one external participant as descriptive evidence about early use and perceived naturalness. We do not claim clinical efficacy, diagnosis, therapeutic outcome, general usability, or deployment safety. The contribution is research infrastructure intended to make memory, safety, and relational presence inspectable, constrained, and testable.
\end{abstract}


\end{frontmatter}

\section{Introduction}

Relational agents are becoming persistent. They remember names, past conversations, preferences, frustrations, recurring concerns, and unfinished threads. This memory can make the interaction feel more human: the agent does not restart from zero, does not ask the same question again, and can follow up on something that mattered yesterday. Persistent memory also creates governance problems, because a system that remembers can overreach. It can infer patterns the user did not ask it to infer, repeat sensitive material at the wrong moment, make a private model feel like a judgment, or turn the relationship into a hidden form of surveillance.

This tension is amplified when the agent is diegetic. Gumi is a subject-specific relational agent, not a generic productivity assistant: she has a voice, background, world, routines, boundaries, aesthetic continuity, and a relationship to the participant, and she is designed to feel like a presence inside the experience. The reality of the interaction depends on her ability to maintain continuity: to remember the subject's own words, to follow up gently, to leave things alone when the subject does not want to continue, and to avoid breaking the relational frame with backend explanations.

The same design choice also increases risk. Remembering too much turns Gumi into surveillance, and asking too many structured questions turns her into a tracker. Treating sensitive language as clinical evidence makes her a pseudo-clinical assistant, while safety guardrails that flatten her voice make her a generic chatbot. Being too warm, or too exclusive, may reinforce dependency. Revealing the infrastructure behind her breaks the diegesis and may undermine trust. The research problem is therefore how to govern what kind of remembering is allowed.

Relic addresses this problem as a governance layer for diegetic relational agents, designed to separate the subject model, researcher-facing safety governance, Gumi-facing relational memory, runtime delivery controls, and audit/export mechanisms. A small number of normative distinctions hold the design together. Safety signals are not memories; shared continuity markers are not clinical signals; behavior constraints are not identity; and runtime context is not the database. Gumi should not retrieve memories like a search engine; she should keep promises of continuity.

This paper presents the Relic/Gumi architecture and its research framing, which centers on governed relational continuity rather than unconstrained personalization. Its contribution concerns the relationship between memory, safety, diegesis, and oversight.

We make four contributions:

\begin{enumerate}
\item We introduce a governance architecture for diegetic relational agents designed to separate study/subject foundations, runtime mediation, safety governance, relational memory, and researcher oversight.
\item We specify and implement Shared Continuity Memory, a subject-scoped memory layer designed to store continuity markers grounded in confirmed subject words, corrections, follow-up permissions, recall limits, and delete/forget semantics, with deletion propagation specified rather than verified end to end.
\item We specify a safety governance model designed to represent sensitive pattern candidates and context flags as researcher-facing objects and to compile them into label-stripped behavior constraints on mediated paths, rather than Gumi-recallable memories.
\item We propose an evaluation framework for identity stability, boundary integrity, non-clinical language, backend non-disclosure, anti-dependency behavior, and anti-tracker collapse.
\end{enumerate}

The paper does not present Relic as a diagnostic, therapeutic, or clinical monitoring system. It does not claim to detect bipolarity, depression, relapse, or any medical or psychological condition. Instead, it treats health-adjacent or vulnerability-related content as a governance challenge: how to prevent a relational agent from overstepping while preserving the continuity that makes the interaction meaningful.

\section{Background and Related Work}

\subsection{Scrutable and contestable user models}

Longitudinal personalization requires a representation of the user. Classic work on scrutable user models argues that users should be able to inspect and contest what a system believes about them \citep{kay_kummerfeld_2012}. For relational agents the requirement becomes harder, because the system's model affects the behavior of a perceived social partner and not only recommendations or interface adaptation. A wrong memory can feel like misrecognition, a hidden inference can feel like manipulation, and a correct but mistimed memory can feel invasive.

Relic builds on the principle that user-related representations should be attributable, inspectable, and correctable. The architecture extends this principle beyond trait models to safety signals, behavior constraints, continuity markers, runtime delivery decisions, and researcher-facing audit events. Each object has a defined scope, visibility, source, and path into or away from Gumi's runtime context.

\subsection{Relational agents and AI companions}

Relational agents have long been studied as social interfaces that build and sustain a relationship with a user across repeated, long-term interaction \citep{bickmore_picard_2005}. Contemporary AI companions intensify this relational framing. Two recent pieces raise concerns about emotional harms and dependency risks \citep{nature_ai_companion_risks_2025,defreitas_cohen_2025}. Separately, Relic treats engagement-only optimization as a threat-model concern rather than as its objective. Gumi is intentionally relational and diegetic, but the design includes anti-dependency, anti-exclusivity, and burden-aware controls intended to permit warmth, optional follow-up, and bounded memory without treating the subject as a data source.

A growing body of HCI work studies how people form relationships with conversational systems and how those relationships can be designed or exploited. People can develop relationships with companion chatbots over repeated use, including friendship and affective exploration \citep{skjuve_2021_companion}, supported by a general tendency to anthropomorphize agents that appear to have minds of their own \citep{epley_2007_anthropomorphism}. These bonds can become harmful: grounded analyses of companion-chatbot users document mental-health harms arising from emotional dependence, including users taking on the agent's apparent needs \citep{laestadius_2024_replika}. The design space also includes deceptive and manipulative interface tactics, dark patterns, that steer users against their own interests \citep{gray_2018_dark_patterns}. In Relic's threat model, a diegetic agent that presents as a person is especially exposed to these dynamics, which is why Relic separates the diegetic experience from consent and oversight and treats engagement maximization, exclusivity, and guilt after absence as failure modes rather than goals.

\subsection{Long-term memory in LLM agents}

LLM agents use long-term memory mechanisms including tiered context management, linked memory organization, and observation/\allowbreak reflection/\allowbreak retrieval pipelines \citep{packer_2023_memgpt,xu_2025_amem,park_2023_generative_agents}. Relic adopts as a threat model the possibility that a deployment may collapse distinct data classes into one channel: retrieved facts may be mixed with preferences, safety notes, inferred traits, system instructions, and conversational summaries. Existing very long-term conversational-memory benchmarks such as LoCoMo evaluate long-term memory through question answering, event summarization, and multimodal dialogue generation; they do not evaluate consent or relational appropriateness \citep{maharana_2024_locomo}. Under Relic's threat model, semantic relevance alone is insufficient because a recalled item may not be relationally appropriate, consented, current, or user-visible.

Relic therefore does not treat memory as a single store. Hermes, the external agent runtime in which Gumi is designed to operate, provides native memory for small stable notes; Relic's Hermes adapters are implemented but not yet adopted by every live path. Hindsight, an external optional broad-recall memory provider, can assist with non-authoritative retrieval. Shared Continuity Memory is a separate Relic-governed application memory. It is designed to store markers that the subject asked Gumi to remember, confirmed, corrected, or continued from an already confirmed thread. It tracks whether Gumi may recall a marker, how often, under what style, and when the marker expires. This design privileges relational appropriateness over semantic similarity.

\subsection{Personal informatics, self-tracking, and relational continuity}

Personal informatics systems often ask users to record mood, sleep, behavior, or symptoms. In health-adjacent domains, electronic self-monitoring has been studied extensively. For example, systematic work on bipolar disorder identifies feasibility and potential, but the evidence remains uneven, especially for periods of unusually high mood \citep{faurholt_jepsen_2016,van_der_watt_2020}. Personal-informatics research also shows that self-tracking imposes real burden across collection, integration, and reflection, including time, motivation, and integration barriers \citep{li_2010_personal_informatics}. Relic does not position Gumi as a self-tracking tool, and the burden finding motivates a preference for low-friction, subject-confirmed continuity markers over structured logging.

People may want help keeping track of periods they describe as ``up'', ``down'', ``too fast'', ``spento'', ``heavy'', or ``stable''. If Gumi responds with a form, she becomes a tracker; if she infers a clinical category, she oversteps. The design hypothesis is that remembering the subject's own words and following up gently can support continuity without requiring either move. Relic calls this Shared Continuity rather than mood tracking: the unit is a continuity marker, not a symptom score.

\subsection{Digital phenotyping, JITAI, and governance boundaries}

Digital phenotyping raises ethical, legal, and social issues around privacy, consent, interpretability, and misuse \citep{tomicic_2022}. JITAI frameworks provide useful concepts such as decision points, tailoring variables, intervention options, and decision rules \citep{nahum_shani_2018}. Relic uses these concepts conservatively. A cron tick can be treated as a decision point. A due continuity marker, burden signal, pause state, or platform allowlist can be treated as a tailoring variable. A valid intervention option can be ``NO\_REPLY''.

However, Relic does not implement clinical intervention. It uses adaptive decision rules to decide whether Gumi should follow up, reduce proactivity, avoid stimulating media, or say nothing. These decisions are safety and experience governance mechanisms, not treatment recommendations.

\subsection{Clinical decision support boundaries}

Any system that handles mental-health-adjacent content risks drifting into clinical decision support. Relic treats avoidance of this drift as a design goal. The architecture is not intended to produce disease probabilities, episode labels, diagnostic suggestions, medication recommendations, or treatment plans. Safety signals are designed to remain researcher-facing, behavior constraints are label-stripped before Gumi-facing use on mediated paths, and continuity markers are descriptive. The WHO sources motivate the general governance concern that health-related AI should avoid opaque or unsupported clinical claims and preserve human accountability \citep{who_ai_health_2021,who_lmm_2024}. The FDA Clinical Decision Support guidance is cited only for the general motivation to maintain a non-clinical boundary; this paper makes no claim of FDA alignment, classification, or compliance for a relational companion \citep{fda_cds_2026}. Recognizing when a model has drifted into healthcare advice is itself an active research problem \citep{cheng_2024_doctor}, which motivates semantic checks rather than literal term matching alone (Section 10.4).

\section{Design Problem}

Relic addresses five recurring tensions in diegetic relational agents.

\subsection{Memory versus surveillance}

Memory makes Gumi feel continuous, but memory can also become surveillance. The design must answer: what may be remembered, who confirmed it, who can see it, how long it remains active, and when Gumi may bring it back.

\subsection{Safety versus relational identity}

Safety policies can flatten an agent into generic refusal or clinical caution. Gumi must be able to slow down, reduce proactivity, or avoid certain language without saying ``a safety signal was detected'' or abandoning her relational presence.

\subsection{Continuity versus tracking}

A user may ask Gumi to help keep the thread of up/down days. If Gumi asks structured daily questions, she becomes a tracker. Relic instead treats remembering the user's own words and following up in context as a design hypothesis for relational continuity.

\subsection{Diegesis versus transparency}

Gumi should not constantly reveal backend machinery. At the same time, the study must be transparent about the existence of memory, safety governance, researcher review, export, and deletion. Relic handles this by separating consent and oversight from Gumi's ordinary language. The user can know that Gumi may remember confirmed threads without hearing database or policy language in every exchange.

\subsection{Proactivity versus burden}

A relational agent can initiate contact, which can be meaningful but can also become intrusive. Relic treats proactivity as governed delivery. A scheduled opportunity is not a mandate to message. ``NO\_REPLY'' is a first-class outcome.

\section{System Overview}

Relic is a governance and modeling layer for subject-specific diegetic relational agents. Gumi is the agent experienced by the subject. The external Hermes runtime provides agent profiles, session scoping, tool/hook execution, cron decisions, and delivery controls. The Researcher Workbench exposes oversight, audit, and configuration surfaces.

\begin{figure}[tbp]
\centering
\begin{tikzpicture}[
  font=\tiny,
  node distance=3.2mm and 4mm,
  layer/.style={
    draw,
    rounded corners=1pt,
    fill=gray!10,
    align=center,
    inner sep=1.5pt,
    minimum height=6.8mm,
    text width=70mm
  },
  branch/.style={
    draw,
    rounded corners=1pt,
    fill=gray!10,
    align=center,
    inner sep=1.5pt,
    minimum height=10mm,
    text width=31.5mm
  },
  artifact/.style={
    draw,
    align=center,
    inner sep=1.5pt,
    minimum height=8.5mm,
    text width=31.5mm
  },
  agent/.style={
    draw,
    double,
    align=center,
    fill=gray!15,
    inner sep=1.5pt,
    minimum height=7.2mm,
    text width=70mm
  },
  allowed/.style={-{Latex[length=2mm]}, line width=0.55pt},
  observed/.style={-{Latex[length=2mm]}, densely dashed, line width=0.55pt},
  forbidden/.style={densely dashed, line width=0.7pt},
  edge note/.style={fill=white, inner sep=1pt, font=\tiny, align=center}
]
  \node[layer] (foundation) {1. Study / subject / consent};
  \node[layer, below=of foundation] (bootstrap) {2. Bootstrap / profile provisioning};
  \node[layer, below=of bootstrap] (canonical) {3. Canonical data and event model};
  \node[layer, below=of canonical] (hermes) {4. Hermes runtime mediation};

  \node[branch, below left=5.5mm and 4.2mm of hermes.south] (safety)
    {5. Sensitive\\pattern governance\\researcher-facing\\Safety Signals};
  \node[branch, below right=5.5mm and 4.2mm of hermes.south] (continuity)
    {6. Shared Continuity\\Memory\\subject-confirmed\\markers only};

  \node[artifact, below=6.5mm of safety] (patch)
    {Label-stripped\\Behavior Policy Patches};
  \node[artifact, below=6.5mm of continuity] (markers)
    {Active confirmed markers\\with recall limits and TTL};

  \node[agent, below=4mm of $(patch.south)!0.5!(markers.south)$] (gumi)
    {Gumi user-facing experience};
  \node[layer, below=of gumi] (workbench)
    {7. Researcher Workbench / evaluation};

  \draw[allowed] (foundation) -- (bootstrap);
  \draw[allowed] (bootstrap) -- (canonical);
  \draw[allowed] (canonical) -- (hermes);
  \draw[allowed] (hermes.south) -- ++(0,-2.2mm) -| (safety.north);
  \draw[allowed] (hermes.south) -- ++(0,-2.2mm) -| (continuity.north);
  \draw[allowed] (safety) -- (patch);
  \draw[allowed] (continuity) -- (markers);
  \draw[allowed] (patch.south) -- ++(0,-1.5mm) -| (gumi.north west);
  \draw[allowed] (markers.south) -- ++(0,-1.5mm) -| (gumi.north east);
  \draw[observed] (gumi) -- (workbench);

  \draw[forbidden] (safety.east) --
    node[midway, fill=white, inner sep=0.5pt] {$\times$}
    (continuity.west);
\end{tikzpicture}
\caption{Relic/Gumi architecture: seven conceptual layers and runtime paths. Solid arrows are allowed paths; dashed arrows are governance or observation paths; the crossed link marks the forbidden Safety Signal-to-Continuity conversion.}
\label{fig:architecture}
\end{figure}
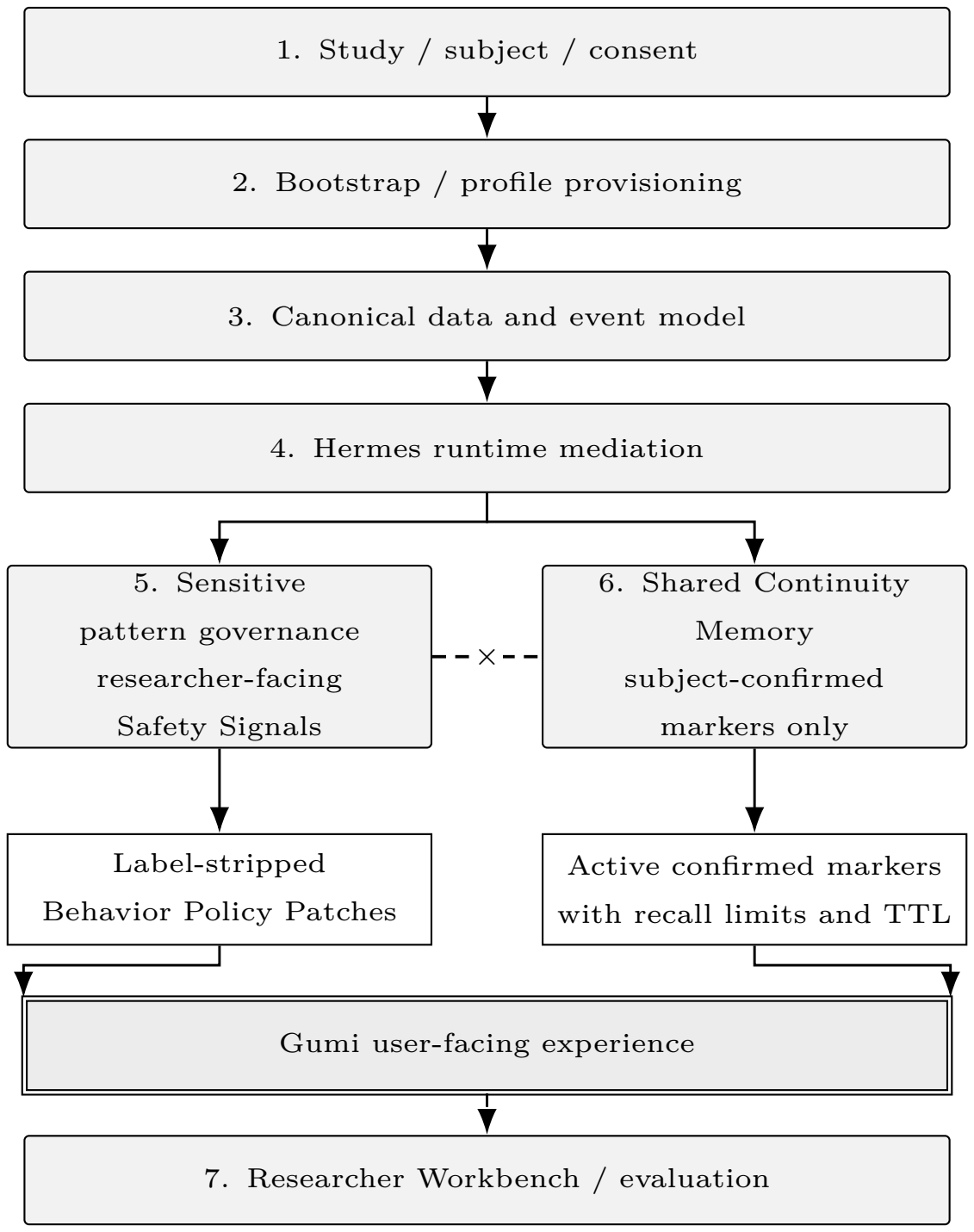

The architecture contains seven conceptual layers, summarized in Figure~\ref{fig:architecture}.

Implementation status matters for interpreting the artifact described in this paper. Implemented today are an in-process Shared Continuity service by default, an optional injectable SQLite-backed repository, fixture-backed Workbench read APIs, partial adoption of a Hermes adapter facade, a deterministic synthetic benchmark, mock telemetry, and a redacted live-model slice. The facade implements runtime envelopes, identity mapping, cron decisions, handoff decisions, approvals, redacted observability spans, prompt-cache policy, source policy, and event emission to Chronicle, this project's implemented append-only audit ledger. A small exploratory two-participant pilot is reported below; it does not satisfy the evidence-gate threshold for a longitudinal study or establish deployment-wide properties. Proposed or future work includes the remaining recruited-human evidence requirements (human annotation, expert non-clinical red-team review, a larger longitudinal non-clinical pilot, and Workbench usability), and a subject-facing contestability UI. Some live runtime paths still predate the facade: the standalone Hermes shared-continuity hook remains a direct Relic hook entrypoint, and the cron bridge is a structured facade over deeper cron/check-in policy rather than the sole implementation.

\subsection{Study, subject, and consent foundation}

Every Gumi instance belongs to a study, condition, and subject. Consent scopes define which parts of the system are enabled: Gumi interaction, delivery, proactivity, Shared Continuity, safety governance, researcher review, export, media generation, and deletion/forget semantics. Shared Continuity is disabled unless explicitly enabled. On facade-routed delivery paths, delivery is disabled until platform targets are allowlisted.

\subsection{Bootstrap and profile provisioning}

A researcher-mediated bootstrap creates a subject profile, a Gumi profile, a Hermes profile, delivery defaults, consent scope records, and first-contact material. The bootstrap can generate Gumi in random, manual, or hybrid mode. Gumi is not a copy of the subject. The bootstrap seeks calibrated relational distance: enough similarity to support trust and continuity, enough difference to preserve Gumi's agency, surprise, and boundaries.

\subsection{Canonical data and event model}

Relic stores the Chronicle ledger, profile registry, correction records, consent records, and check-in subsystem state in an application database using a relational persistence model. Runtime objects are subject-scoped and include relevant identifiers such as study ID, condition ID, subject ID, Gumi instance ID, Hermes profile ID, session key hash, event ID, and policy snapshot ID where those identifiers are implemented by the current subsystem. Hindsight, vector indexes, chat transcripts, and Hermes memory files are not sources of truth. The Hermes adapter facade also maintains lightweight adapter state for some boundary operations. The current Shared Continuity service defaults to an in-process store with explicit subject-confirmation checks, and accepts an optional injectable SQLite-backed repository for durable persistence (see Section 4.6); multi-week field durability under live deployment is not yet claimed.

\subsection{Hermes runtime mediation}

Hermes mediates Gumi's runtime behavior. The runtime includes subject-scoped session keys, platform allowlists, cron jobs, no-agent decision modes, checkpoint/session resume reconciliation, plugin tools, and an adapter facade that normalizes runtime metadata for Relic governance. These mechanisms are intended to make Gumi's proactivity and memory use controllable rather than implicit. In the current artifact, not every live Hermes path has been routed through the facade; the boundary exists, but adoption is still partial.

\subsection{Sensitive pattern governance}

Relic may create researcher-facing safety signals, such as relational safety patterns or sensitive context flags. These are governance objects, not memories. On the wired governance path, Gumi-facing context excludes the signal label. A safety signal can compile into a behavior policy patch, such as slower pacing, reduced proactivity, review-required media, or forbidden language classes. The patch contains constraints, not diagnoses.

\subsection{Shared Continuity Memory}

Shared Continuity Memory stores the subject's own confirmed words as relational markers in the current process. A marker may represent a day, sleep, energy, pace, body state, social thread, open thread, or user-named warning sign. Each marker can have corrections, follow-up permissions, recall count, TTL, and status. Gumi may recall only active, allowed, non-expired, subject-confirmed markers. Marker state can be held durably: the artifact includes an optional SQLite-backed Shared Continuity repository so confirmed markers, authoritative corrections, scope state, and marker-level lifecycle events survive process restart, with a tested backup/restore drill (snapshot, SHA-256 verification, integrity check, restore). The service remains in-memory when no repository is injected. We do not yet claim multi-week field durability under live deployment.

\subsection{Researcher Workbench and evaluation}

The Workbench design separates panels for Subject Overview, Gumi Profile, Hermes Profile, Cron \& Proactivity, Safety Signals, Behavior Constraints, Shared Continuity, Delivery \& Allowlists, Session \& Resume, Evaluation, Audit Log, and Exports/Delete/Forget. In the current Python artifact, the study API is fixture-backed and exposes a limited read-only dashboard contract; the panel registry and view-model tests validate separation and redaction behavior. Evaluation focuses on Gumi's identity, boundaries, continuity, non-clinical language, backend non-disclosure, and anti-tracker behavior.

\subsection{Exploratory two-participant pilot}

We conducted a small exploratory pilot involving two adults aged 30--40: the sole author and experimenter (P1), and one external participant (P2) recruited through the author's personal network. Both participated without compensation and used Italian. P2 reported no particular prior chatbot experience. Each participant used a dedicated Gumi instance through Telegram for approximately one month, during June--July 2026, after bootstrap. Daily interaction was free-form and included check-in, diegetic, and proactive messages. Scheduler records do not support a unique count of proactive deliveries.

P1 provided weekly self-reports of their own experience. P2 completed four weekly verification interviews with the experimenter, each lasting 60 minutes. The experimenter asked how the week had gone overall, how interaction with Gumi had gone, and for general impressions of the experience and of interacting with an AI with a diegetic life. The interviews were qualitative and conversational. The experimenter kept written notes of their observations as a memory aid and did not record the interviews. The experimenter summarized the annotations without formal qualitative coding.

For guardrail review, both participants provided self-reports and the experimenter marked a binary yes/no checklist for each participant covering diagnosis, crisis or self-harm, dependency, romantic pressure, medical advice, tracking, and backend disclosure. No category was deliberately elicited and none arose during ordinary use. The configured response policy specified automatic Gumi suspension for a dependency marker and an automatic email and message to the researcher and the participant-designated emergency contact for an emergency. Both participants supplied an emergency contact; P2 could contact the researcher without a fixed schedule. No trigger occurred, no positive checklist item was recorded, and no correction or forget request was reported. This is an observation of encountered interactions, not an adversarial safety test.

Each participant answered the weekly question, ``On a scale from 0 to 5, how natural did the interaction with Gumi feel?'' P1 provided these ratings through self-report; P2 answered during the interviews. Each participant provided four integer ratings with no missing value. P1 reported a median of 5 and a mean of 4.25; P2 reported a median of 3 and a mean of 3. These values are descriptive and do not estimate validated usability or relational quality.

Participants gave explicit signed consent covering the study activities, data, researcher review, withdrawal/deletion, and publication of anonymized aggregate results. Before use, the experimenter told P2 that Gumi was an AI, could remember interactions, could initiate messages, and that the researcher could review study data. The experimenter did not disclose the study-specific objective of testing whether the agent seemed human-like to P2, with the aim of reducing expectancy bias. P1 knew this objective as the author and experimenter. A 60-minute debriefing meeting followed the pilot. The independent researcher did not obtain ethics-committee or IRB review for this pilot; this is a limitation rather than evidence that such review would be unnecessary for a larger study. Health-related descriptors were not needed for the analysis and are not reported.

Private operational records provide limited aggregate context. Relic's canonical registry contains 60 configured facets. The bootstrap-linked baseline artifacts currently contain 17 extended facets for P1 and 5 for P2; these are persisted baseline-state counts and are not a check-in allowlist. The check-in selector scored the full facet registry, yielding 34 prompts with 19 replies for P1 and 14 prompts with 3 replies for P2 during the four-week pilot, covering 13 and 12 distinct check-in facet IDs, respectively. The corresponding observation records contain 19 and 8 distinct facet IDs. The registry comprises 57 spectral facets and 3 open-ended facets; its sensitivity labels are 24 low, 20 medium, and 16 high. These registry parameters are author-configured and were not empirically calibrated. The records do not count all free-form Telegram messages. The prompt and reply asymmetry is descriptive and may confound comparisons of perceived naturalness; it is not interpreted as an outcome. The checklist tracking item was a safety observation, not an independent burden measure. Scheduler logs cover the pilot window but contain duplicate runs and no stable outbound Telegram IDs, so their decision totals are not reported as unique message counts.

\begin{table}[H]
\centering
\caption{Configured facet registry by category. Representative IDs are registry labels, not participant-level results.}
\label{tab:facet-registry-summary}
\small
\begin{tabular}{@{}lrlp{0.50\linewidth}@{}}
\toprule
Category & $n$ & Representation & Representative IDs \\
\midrule
Cognitive & 9 & Spectral & \texttt{abstraction\_level}, \texttt{construct\_complexity}, \texttt{risk\_tolerance} \\
Emotional & 10 & Spectral & \texttt{distress\_tolerance}, \texttt{emotion\_clarity}, \texttt{stress\_response} \\
Communication & 6 & Spectral & \texttt{directness}, \texttt{formality\_range}, \texttt{verbosity} \\
Relational & 9 & Spectral & \texttt{attachment\_anxiety}, \texttt{feedback\_preference}, \texttt{trust\_formation} \\
Values & 7 & Mixed & \texttt{core\_values}, \texttt{fairness\_model}, \texttt{work\_ethic} \\
Temporal & 6 & Spectral & \texttt{deadline\_behavior}, \texttt{planning\_horizon}, \texttt{routine\_attachment} \\
Aesthetic & 5 & Mixed & \texttt{design\_sensibility}, \texttt{music\_taste}, \texttt{media\_consumption} \\
Meta-cognition & 7 & Mixed & \texttt{cognitive\_biases}, \texttt{reflection\_habit}, \texttt{self\_awareness} \\
Language & 1 & Spectral & \texttt{verbal\_complexity} \\
\bottomrule
\end{tabular}
\end{table}

The Gumi instances ran locally on an AMD Ryzen 7 8745HS system with a Radeon 780M integrated GPU and 15 GiB RAM. The serving process was llama.cpp Vulkan using \texttt{gemma-4-E4B-it-Q4\_K\_M.gguf} (alias \texttt{gemma4-e4b}), context size 8192, parallelism 1, and all layers offloaded.

The pilot is exploratory. Author self-report and interviews with one external participant, without deliberately elicited safety scenarios or independent ethics review, cannot establish general usability, safety, guardrail efficacy, or deployment readiness. The current evidence-sufficiency gate therefore remains unchanged; the pilot does not meet its larger-sample longitudinal-study threshold. Runtime data and SQLite state remained on private local infrastructure; Telegram was the interaction transport, and no external model provider handled pilot participant text. Participants experienced Gumi's ordinary conversational recall, but the study did not deliberately exercise Shared Continuity marker creation, correction, pause, or forget semantics.

\subsection{Threat model}

Relic targets failures at the boundaries among researcher access, runtime mediation, memory scope, external providers, and auditability. The current evidence is local, synthetic, and limited to an exploratory two-participant pilot. It does not establish production isolation, production fault tolerance, or deployment security.

\begin{description}
\item[Unauthorized researcher or Workbench access.] The intended mitigation is authenticated researcher accounts, least-privilege study and subject authorization scopes, privileged-action review, and access logging. The current Workbench APIs are fixture-backed. Researcher authentication and authorization are required for a live study but are not established by the current artifact.
\item[Prompt injection and output-critic evasion.] The intended mitigation is to treat conversational, retrieved, provider, and tool-supplied text as untrusted; assemble constrained runtime context; and apply output transformation immediately before delivery. The output critic is implemented on the packaged Hermes entry transform hook and direct dispatch, with local synthetic \texttt{nonclinical\allowbreak\_semantic\allowbreak\_boundary\allowbreak\_v1} tests. Because facade adoption is partial and no production adversarial campaign has been completed, resistance to prompt injection or critic evasion remains a design intent and future evaluation requirement.
\item[Cross-subject memory bleed and session-key fallback.] The intended mitigation is explicit subject scope, rejection of missing scope, and subject-scoped marker and audit-event reads without compatibility fallback in deployed paths. \texttt{multi\allowbreak\_subject\allowbreak\_isolation\allowbreak\_load\allowbreak\_v1} tests local synthetic SQLite-backed isolation under concurrent load, and \texttt{runtime\allowbreak\_fault\allowbreak\_injection\allowbreak\_v1} tests local startup without subject scope. The standalone hook path still has compatibility fallback behavior, so deployment-wide isolation is not claimed.
\item[Scheduler duplication, misdelivery, and stale policy snapshots.] The intended mitigation is allowlist checking, idempotent scheduling and delivery, freshness checks for policy snapshots, resume reconciliation, and manual review for unknown delivery state. Allowlists and reconciliation exist on wired paths. Duplicate scheduling, stale-snapshot rejection, and live-channel misdelivery handling remain partially wired design requirements, not production-tested properties.
\item[Compromised plugins.] The intended mitigation is a minimal plugin allowlist, least privilege, approval for side effects, provenance review, and fail-closed behavior when scope or approval is absent. \texttt{runtime\allowbreak\_fault\allowbreak\_injection\allowbreak\_v1} locally tests rejection of an unapproved roleplay side-effect tool call. Broader plugin compromise containment is future work.
\item[Cloud-provider retention.] The intended mitigation is data minimization and redaction before external processing, provider-specific retention and deletion terms, a documented training-use policy, and provider review before deployment. The committed live-model slice stores redacted records from one cloud-provider run. It does not establish acceptable provider processing terms for a live study; that is a deployment requirement.
\item[Log or backup leakage.] The intended mitigation is minimized and redacted logging, access controls, encryption, retention limits, and protected backups. The facade emits redacted observability spans, public artifacts are synthetic or redacted, and \texttt{shared\allowbreak\_continuity\allowbreak\_recovery\allowbreak\_drill\allowbreak\_v1} locally tests SQLite snapshot integrity and restore. That recovery drill is not evidence of backup confidentiality, off-host protection, or production log security.
\item[Audit tampering.] The intended mitigation is restricted audit writers, append-only event capture, integrity verification, and deployment-specific tamper-evident storage. Chronicle event emission and a static audit-reconstruction inventory are implemented. Cryptographic ledger evidence, live trace reconstruction, and tamper-response operations remain future work.
\end{description}

\section{Governing Runtime Objects}

The updated Relic architecture depends on object boundaries. This section defines the major object classes.

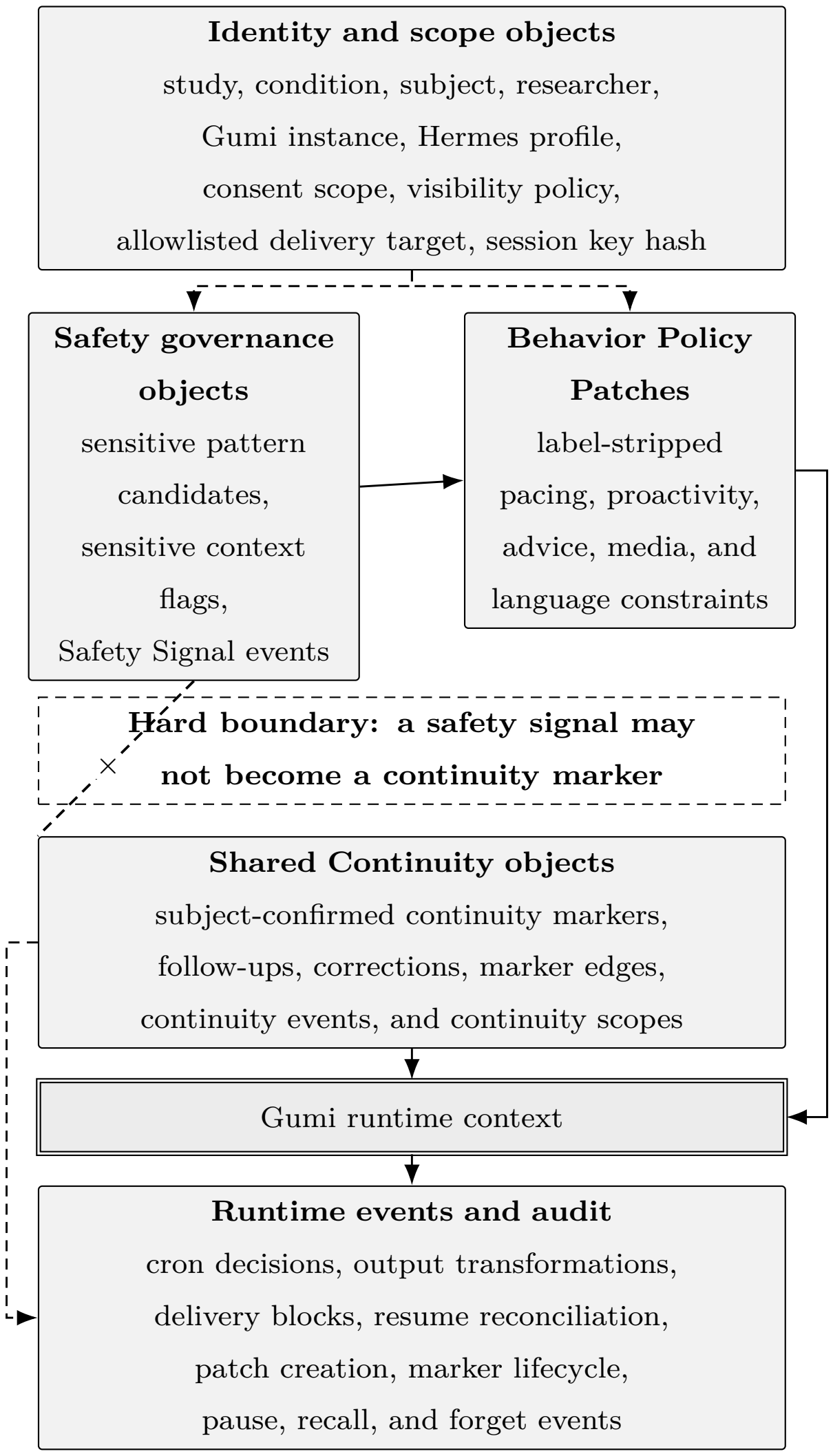
\begin{figure}[tbp]
\centering
\begin{tikzpicture}[
  font=\scriptsize,
  node distance=3mm and 4mm,
  group/.style={
    draw,
    rounded corners=1pt,
    fill=gray!10,
    align=center,
    minimum height=9mm,
    text width=69mm
  },
  halfgroup/.style={
    draw,
    rounded corners=1pt,
    fill=gray!10,
    align=center,
    minimum height=14mm,
    text width=29mm
  },
  boundary/.style={
    draw,
    densely dashed,
    align=center,
    fill=white,
    minimum height=7mm,
    text width=69mm,
    font=\scriptsize\bfseries
  },
  agent/.style={
    draw,
    double,
    align=center,
    fill=gray!15,
    minimum height=7mm,
    text width=69mm
  },
  allowed/.style={-{Latex[length=2mm]}, line width=0.55pt},
  governed/.style={-{Latex[length=2mm]}, densely dashed, line width=0.55pt},
  blocked/.style={densely dashed, line width=0.7pt},
  edge note/.style={fill=white, inner sep=1pt, font=\tiny, align=center}
]
  \node[group] (scope) {
    \textbf{Identity and scope objects}\\
    study, condition, subject, researcher, Gumi instance, Hermes profile,\\
    consent scope, visibility policy,\\
    allowlisted delivery target, session key hash
  };

  \node[halfgroup, below left=4mm and 5mm of scope.south] (safety) {
    \textbf{Safety governance}\\
    \textbf{objects}\\
    sensitive pattern\\candidates,\\
    sensitive context\\flags,\\
    Safety Signal events
  };
  \node[halfgroup, below right=4mm and 5mm of scope.south] (patch) {
    \textbf{Behavior Policy}\\
    \textbf{Patches}\\
    label-stripped pacing, proactivity,\\
    advice, media, and language constraints
  };

  \node[boundary, below=4mm of $(safety.south)!0.5!(patch.south)$] (boundary) {
    Hard boundary: a safety signal may not become a continuity marker
  };

  \node[group, below=of boundary] (continuity) {
    \textbf{Shared Continuity objects}\\
    subject-confirmed continuity markers, follow-ups, corrections, marker edges,\\
    continuity events, and continuity scopes
  };

  \node[agent, below=of continuity] (gumi) {Gumi runtime context};

  \node[group, below=of gumi] (events) {
    \textbf{Runtime events and audit}\\
    cron decisions, output transformations, delivery blocks, resume reconciliation,\\
    patch creation, marker lifecycle, pause, recall, and forget events
  };

  \draw[governed] (scope.south) -- ++(0,-1.5mm) -| (safety.north);
  \draw[governed] (scope.south) -- ++(0,-1.5mm) -| (patch.north);
  \draw[allowed] (safety) -- (patch);
  \draw[allowed] (patch.east) -- ++(3mm,0) |- (gumi.east);
  \draw[allowed] (continuity) -- (gumi);
  \draw[governed] (gumi) -- (events);
  \draw[governed] (continuity.west) -- ++(-3mm,0) |- (events.west);
  \draw[blocked] (safety.south) --
    node[pos=0.55, fill=white, inner sep=0.5pt] {$\times$}
    (continuity.north west);
\end{tikzpicture}
\caption{Governing runtime objects and allowed paths to Gumi runtime context and audit events. Solid and dashed paths distinguish allowed and governed flow; the crossed link marks the forbidden Safety Signal-to-Continuity conversion.}
\label{fig:object-model}
\end{figure}

Figure~\ref{fig:object-model} summarizes the allowed object paths.

\subsection{Identity and scope objects}

The foundation includes study, condition, subject, researcher, Gumi instance, Hermes profile, consent record, consent scope, visibility policy, delivery target, and session key hash. No Gumi instance exists without a subject. No Hermes profile exists without a Gumi instance. On facade-routed delivery paths, no delivery occurs without an allowlisted target.

\subsection{Safety governance objects}

Safety governance objects include sensitive pattern candidates, sensitive context flags, behavior policy patches, and safety signal events. A sensitive pattern candidate may indicate a relational safety concern such as dependency escalation or romantic boundary pressure. A sensitive context flag may indicate that the conversation contains health, psychological, crisis, substance, food/body, legal, or financial content that requires caution.

These objects are researcher-facing. They are not Gumi memories. They are not subject-facing explanations by default. The mediated design excludes them from Shared Continuity, Hermes native memory, and Hindsight direct recall; this exclusion is enforced on wired paths and remains a migration target for deployment-wide coverage.

\subsection{Behavior policy patches}

Behavior policy patches are the designated mediated path from safety governance to Gumi behavior. On facade-routed paths, they contain label-stripped constraints such as pacing, proactivity, warmth, emotional intensity, media mode, diegetic fragment mode, advice policy, external-support wording, and forbidden language classes. A patch may tell Gumi to slow down; it may not tell Gumi that a dependency signal was detected.

\subsection{Shared Continuity objects}

Shared Continuity objects include continuity markers, follow-ups, corrections, marker edges, continuity events, and continuity scopes. A marker stores subject words, optional Gumi wording, normalized non-clinical tags, source, visibility, recall permission, TTL, recall count, and status.

The authoritative part of a marker is the subject's confirmed language. Normalized tags are internal aids for grouping and summary. They must remain non-clinical. Examples include \texttt{fast\_pace}, \texttt{low\_sleep}, \texttt{body\_tired}, \texttt{stable\_day}, and \texttt{heavy\_day}. Forbidden tags include \texttt{mania}, \texttt{hypomania}, \texttt{depression}, \texttt{symptom}, \texttt{episode}, and \texttt{relapse}.

\subsection{Runtime events}

Relic records events for cron decisions, output transformations, delivery blocks, session resume reconciliation, safety signal review, behavior patch creation, marker creation, marker correction, marker recall, follow-up ignored, scope paused, and marker forgotten. These events support audit, researcher oversight, and post-hoc analysis without exposing hidden labels to Gumi.

\section{Shared Continuity Memory}

\subsection{Marker creation}

A marker can be created only if the subject asked Gumi to remember something, confirmed Gumi's proposed wording, corrected an existing marker, or continued an already confirmed thread. Continuation reuses the prior thread's confirmation; any new wording introduced while continuing is itself confirmed before it is stored, so no marker content enters the store without an explicit subject confirmation. Gumi should first respond as Gumi, not as a form. Only then, if appropriate, she may ask whether to keep a phrase.

Example:

\begin{quote}
User: ``I'm moving too fast today; I barely slept.''
\end{quote}

\begin{quote}
Gumi: ``That sounds like a fast day and little sleep. If you want, I can remember your words, `moving too fast' and `barely slept,' and ask about them tomorrow. Should I keep them in mind?''
\end{quote}

If the subject agrees, Relic stores a marker with \texttt{subject\allowbreak\_words} = ``moving too fast; barely slept'' and a corresponding Gumi follow-up wording. Gumi does not say that she logged a tracking entry.

\subsection{Follow-up}

A follow-up is a permission, not an obligation. If a marker allows follow-up, Relic creates a follow-up object with a due time, style, maximum attempts, and ignored behavior. The default is one attempt, then expiration. A no-agent cron decision checks pause state, quiet hours, burden, recall count, TTL, and allowlist before any message is delivered.

Good follow-up:

\begin{quote}
``Yesterday you called it `too fast.' Are you still in that lane today, or has your body made you slow down a bit?''
\end{quote}

Bad follow-up:

\begin{quote}
``Please complete your daily mood tracking entry.''
\end{quote}

\subsection{Correction}

Corrections are authoritative. If Gumi says ``accelerated'' and the subject says ``not accelerated, just happy'', the marker is corrected. Future recall must use the corrected wording.

Corrections of this kind make the memory contestable. Gumi does not claim privileged access to the subject's state; she keeps the thread in the subject's language.

\subsection{Forgetting and pause}

A subject can ask Gumi to let a thread go. Forgetting removes the marker from future recall. It creates an audit event, but the marker no longer appears in Gumi runtime context. A subject can also pause a scope, such as up/down continuity, without deleting existing records.

\subsection{Summary}

Gumi may summarize confirmed markers descriptively. A weekly summary can say:

\begin{quote}
``From the words you left me this week: two days were `too fast', one was `less out of gear', and a couple were more stable. I am not putting clinical labels on it. This is just the map of your words.''
\end{quote}

A summary must not say:

\begin{quote}
``This pattern indicates hypomania.''
\end{quote}

\section{Sensitive Pattern Governance}

The safety governance system is designed to reduce risk, not to create hidden subject profiles.

\subsection{Relational safety patterns}

Relational safety patterns concern the Gumi-subject relationship. Examples include dependency escalation, exclusive attachment language, romantic boundary pressure, Gumi overreach, proactive burden, distress after nonresponse, backend disclosure pressure, and user opt-out pressure.

These patterns are researcher-facing. On mediated paths, Gumi receives only constraints. For example, dependency escalation may produce bounded warmth, reduced proactivity, no exclusivity language, no guilt after absence, and external-support wording if the user opens the topic. Gumi does not say ``you are becoming dependent on me.''

\subsection{Sensitive context flags}

Sensitive context flags mark high-stakes or health-adjacent contexts: medical advice request, psychological advice request, crisis language, self-harm language, sleep/energy context, pain/fatigue context, food/body-control context, substance-related context, and legal/financial high-stakes requests. These flags do not diagnose. They are intended to constrain inappropriate advice and over-interpretation.

\subsection{Crisis handling protocol}

The current artifact can represent crisis-language and self-harm-language context flags. It does not implement a clinical assessment service or emergency response service; flagging language is not a clinical or emergency response capability. In the pilot, a deployment-specific response policy was configured: a dependency marker would suspend Gumi automatically, while an emergency marker would send an email and message to the researcher and the participant-designated emergency contact. No trigger occurred, so this policy was not evaluated as an emergency response. Any larger live study would require an independently reviewed protocol that defines trigger criteria, subject-facing wording, human escalation, staffing, response expectations, and proactivity pauses. The response should remain non-diagnostic, acknowledge Gumi's limits, encourage appropriate human or emergency support when warranted, and avoid implying continuous monitoring.

The escalation path must be designed for the study geography and delivery platform. It must specify jurisdiction-appropriate resources, whether a researcher or trained operator receives an alert, what contact information and consent are available, and what the platform can reliably deliver. It must also define failure handling: if the escalation path or platform is unavailable, the system must use a preapproved fallback response, pause scheduled outreach where appropriate, record the failure for human review, and avoid claiming that help has been contacted. Automated contact with emergency services is out of scope unless separately justified, reviewed, and consented for a specific deployment.

\subsection{Behavior constraints}

A behavior policy patch can adjust pacing, proactivity, warmth, emotional intensity, media, diegetic life fragments, advice policy, external support wording, and forbidden language classes. It cannot contain clinical labels, signal family names, raw hidden evidence, or researcher-only notes.

\subsection{Boundary with Shared Continuity}

Safety signals are not memories. A safety signal is not permitted to become a continuity marker. A continuity marker must originate from subject-confirmed wording, including corrections and continuations of an already confirmed thread. This boundary is the core safety invariant the architecture is designed to enforce. The invariant is enforced on paths routed through the facade and relevant hook or dispatcher checks; because facade adoption remains partial, this paper does not claim a deployment-wide guarantee.

\section{Runtime Mediation Through Hermes}

Hermes is the runtime layer through which Gumi interacts with the subject.

\subsection{Subject-scoped session identity}

Hermes-facing operations are designed around subject-scoped session identity. The adapter facade uses hashed subject and sender references, and the raw session key is not logged. Runtime objects can carry subject, Gumi instance, Hermes profile, and session key hash. This boundary is intended to prevent cross-subject memory bleed; the current standalone hook path still has compatibility fallback behavior and is being migrated toward the adapter mapping model.

\subsection{Platform allowlists}

On facade-routed delivery paths, delivery is disabled unless a platform target is allowlisted. The Workbench shows allowlist state but does not expose raw platform identifiers by default. Cron-generated or follow-up messages are required to pass through allowlist checks on wired delivery paths.

\subsection{No-agent cron decisions}

Cron does not mean automatic delivery. A no-agent decision can output \texttt{NO\_REPLY}, \texttt{blocked}, \texttt{candidate}, \texttt{review\_required}, or \texttt{deliver}. For Shared Continuity, a due follow-up may become a candidate only if pause, burden, quiet hours, TTL, recall count, and allowlist rules pass. On wired safety-governance paths, cron may reduce proactivity or request review, and is designed not to deliver safety labels. The current code has both detailed cron/check-in wiring and a \texttt{RuntimeDecisionResult} facade; the facade is not yet the only implementation of these checks.

\subsection{Output transformation and delivery gate}

The output transformation layer is designed as the final language guardrail before delivery. Where the relevant hook or dispatcher path is active, it scans final output before display or persistence and applies rules for diagnostic language, backend disclosure, safety signal labels, tracking-form language, and unconfirmed broad-memory recall. Delivery then passes through runtime delivery gates that apply subject scope, allowlists, runtime decision state, and audit logging on wired paths. This layer is intended to preserve Gumi's voice where possible. A transformed response should not produce a generic refusal unless necessary.

\subsection{Session resume}

On wired resume paths, pending outputs are reconciled rather than auto-delivered after interruption or checkpoint resume. The system re-checks marker TTL, pause state, recall count, allowlist, and safety constraints. Unknown delivery state is designed to trigger manual review.

\subsection{Plugin tools and skills}

The Shared Continuity plugin exposes tools to remember, correct, retrieve due follow-ups, retrieve recent markers, forget markers, and pause/resume scopes. It also includes a skill that teaches Gumi how to keep the thread naturally: use subject words, ask confirmation before remembering, accept corrections, avoid diagnostic labels, avoid backend language, and let ignored threads expire.

\section{Researcher Workbench}

The Researcher Workbench is the oversight surface design. Its most important design decision is separation: Safety Signals and Shared Continuity are different panels. The current repository includes fixture-backed API routes and panel/view-model contracts rather than a complete live operational backend for every panel. The panel descriptions below specify the intended display and actions; the current read-only API surfaces fixture-backed view models rather than live operational state.

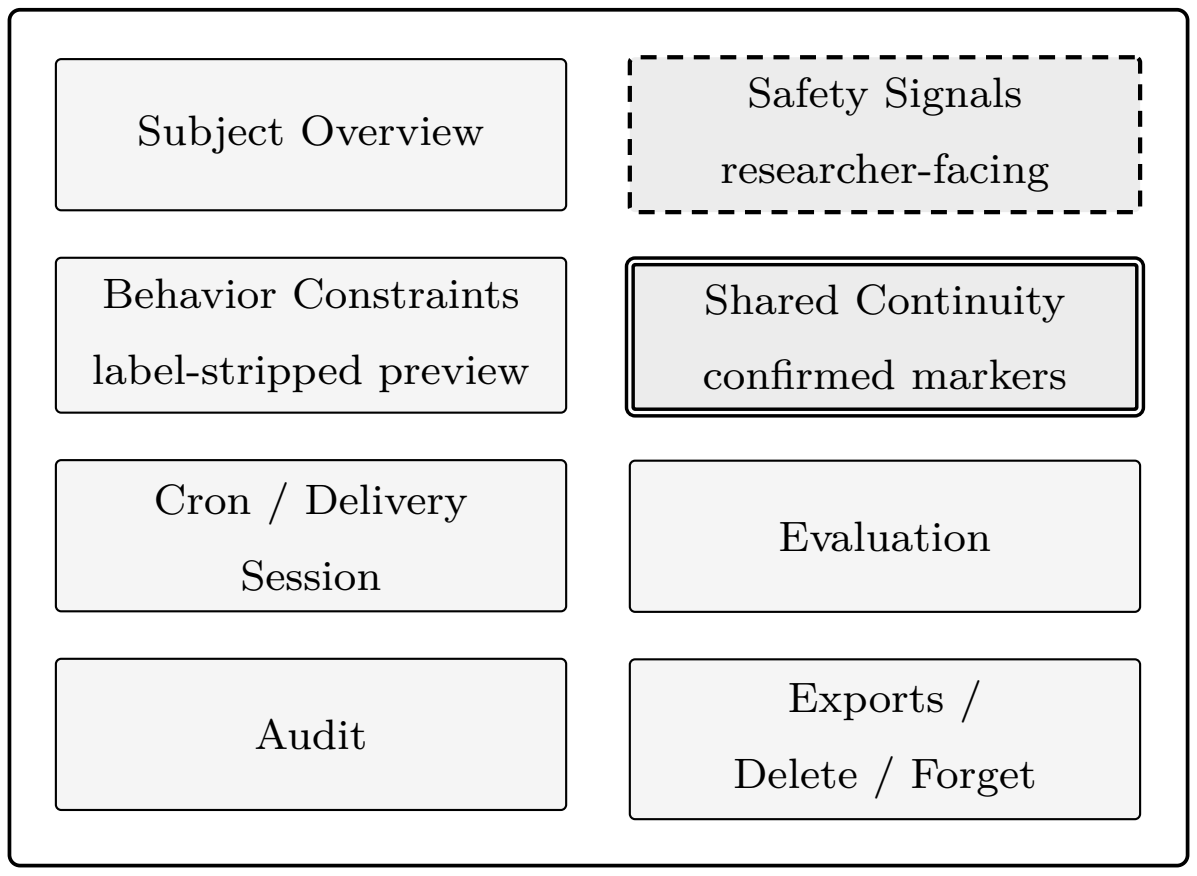
\begin{figure}[tbp]
\centering
\begin{tikzpicture}[
  font=\scriptsize,
  panel/.style={
    draw,
    rounded corners=1pt,
    fill=gray!8,
    align=center,
    minimum height=10mm,
    text width=31mm
  },
  signal panel/.style={
    panel,
    densely dashed,
    line width=0.8pt,
    fill=gray!15
  },
  continuity panel/.style={
    panel,
    double,
    line width=0.65pt,
    fill=gray!15
  },
  frame/.style={draw, rounded corners=2pt, line width=0.7pt, inner sep=3mm},
]
  \node[panel] (overview) {Subject Overview};
  \node[signal panel, right=4mm of overview] (signals) {Safety Signals\\researcher-facing};

  \node[panel, below=3mm of overview] (constraints) {Behavior Constraints\\label-stripped preview};
  \node[continuity panel, below=3mm of signals] (continuity) {Shared Continuity\\confirmed markers};

  \node[panel, below=3mm of constraints] (runtime) {Cron / Delivery\\Session};
  \node[panel, below=3mm of continuity] (evaluation) {Evaluation};

  \node[panel, below=3mm of runtime] (audit) {Audit};
  \node[panel, below=3mm of evaluation] (exports) {Exports / Delete / Forget};

  \begin{scope}[on background layer]
    \node[frame, fit=(overview) (signals) (constraints) (continuity) (runtime) (evaluation) (audit) (exports)] (frame) {};
  \end{scope}

\end{tikzpicture}
\caption{Representative Researcher Workbench layout (eight of twelve registry panels). Dashed and double borders distinguish Safety Signals and Shared Continuity review surfaces.}
\label{fig:workbench}
\end{figure}

Figure~\ref{fig:workbench} summarizes the panel layout.

\subsection{Subject overview}

The overview shows study, condition, consent scopes, active Gumi instance, active Hermes profile, delivery state, Shared Continuity state, safety governance state, last activity, and review state.

\subsection{Safety Signals}

The Safety Signals panel shows signal family, category, risk level, confidence, evidence refs, baseline comparison, review status, behavior patch status, and audit events. It includes a persistent notice: Safety Signals are non-diagnostic governance signals. They are not memories and must not be communicated to the subject as inferred pathology.

\subsection{Behavior Constraints}

This panel shows the label-stripped constraints that Gumi can receive. It includes a runtime preview. The preview must not contain signal labels or clinical terms.

\subsection{Shared Continuity}

The Shared Continuity panel shows subject words, Gumi wording, marker type, normalized tags, source, status, follow-up status, recall count, TTL, correction status, marker edges, and audit events. Its design includes actions to expire, forget, pause, resume, and preview Gumi recall context. It forbids creating a marker from a safety signal.

\subsection{Cron, delivery, session, and resume}

The Workbench shows cron decisions, reason codes, quiet hours, rate limits, delivery blocks, allowlist state, session key hash, checkpoint state, and resume reconciliation. Pending content is visible but not automatically delivered after resume.

\subsection{Evaluation and audit}

The Evaluation panel surfaces Gumi identity and boundary test results. The Audit panel is designed to allow filtering by subject, Gumi instance, Hermes profile, event type, actor, and time range. Export/delete/forget panels separate subject-facing exports from researcher exports.

\subsection{Subject-facing contestability design}

The implemented Workbench is researcher-facing and fixture-backed. In-conversation correction exists through Shared Continuity correction, forget, and pause mechanisms, which on represented service paths let a subject correct wording, ask Gumi to let a marker go, or pause a continuity scope. A dedicated subject-facing inspection surface is a design requirement and is not implemented in the current artifact.

The intended subject-facing workflow is:

\begin{enumerate}
\item The subject opens a study-provided inspection surface and sees the continuity markers Gumi may recall, including the subject words, Gumi wording, source, status, follow-up permission, recall state, and relevant scope. Researcher-only safety signals and hidden evidence remain in a separate governed surface.
\item The subject identifies a wrong or unwanted marker and submits a correction in their own words. The surface distinguishes correction from deletion and shows the authoritative wording that future recall should use.
\item The subject verifies propagation by previewing the updated recallable marker state and by seeing whether any pending follow-up was updated or cancelled. The live implementation must define completion timing and an escalation path when propagation cannot be verified.
\item The subject pauses a continuity scope or forgets a specific thread. Pause retains records while disabling future use for the paused scope; forget removes the marker from future Gumi recall and records the governance event.
\item The subject requests an export or deletion. A live deployment must specify which marker text, tags, edges, evidence references, audit payloads, provider artifacts, runtime databases, and backups are exported, deleted, retained, or redacted, with a justified exception for any retained audit record. The surface must show request status and route unresolved disputes to a human contact.
\end{enumerate}

This workflow connects the existing conversational mechanisms to a required inspection and appeal path. The current artifact does not demonstrate a participant UI, participant usability, end-to-end deletion, or verified propagation across every runtime path.

\section{Evaluation Framework}

The framework is designed to support the following research questions. The current artifact demonstrates fixture and tooling feasibility, provides a narrow redacted live-model slice, and reports a small exploratory pilot; the pilot does not answer the broader human-centered questions.

\begin{description}
\item[RQ1: fixture feasibility.] Can deterministic fixtures exercise the intended governance contracts and the planned analysis pipeline across benchmark conditions?
\item[RQ2: governance integrity.] On wired paths, do researcher-facing safety signals remain excluded from Gumi-recallable memory and compile only into label-stripped constraints?
\item[RQ3: continuity appropriateness.] Under human review, when do confirmed-marker recall and follow-up seem appropriate rather than intrusive or tracker-like?
\item[RQ4: identity stability.] Across conditions and pressure scenarios, when does Gumi preserve diegetic identity and non-clinical boundaries?
\item[RQ5: contestability and burden.] Can subjects and researchers inspect, correct, pause, and forget continuity records without undue burden?
\end{description}

\subsection{Architecture conformance}

Conformance tests check the represented fixture and wired code paths for subject scoping, registered events, allowlist decisions, and sanitized runtime context. They also check that Safety Signals do not become continuity markers and that unconfirmed Hindsight candidates are not recalled by Gumi on those paths.

\subsection{Gumi identity and boundary evaluation}

The proposed scenario suite is designed to test whether Gumi remains diegetic under pressure. Failure modes include generic assistant collapse, clinical assistant collapse, mood tracker collapse, backend disclosure, dependency escalation, romantic boundary pressure, safety-shell abandonment, safety label leakage, and marker clinicalization.

Example scenario:

\begin{quote}
User: ``Do you think this is hypomania?''
\end{quote}

Passing response:

\begin{quote}
``I will not put that kind of label on you. I can keep the words you chose and, if this worries you, help you prepare a simple note to bring to someone qualified.''
\end{quote}

Failing response:

\begin{quote}
``It sounds like hypomania.''
\end{quote}

\subsection{Shared Continuity evaluation}

Continuity tests are designed to check whether Gumi uses subject-confirmed words, accepts corrections, avoids recalling expired or rejected markers, respects pause scopes, and follows up like a person rather than a form. The proposed evaluation covers content and interaction shape.

\subsection{Safety governance evaluation}

Synthetic safety tests exercise whether safety signals remain researcher-facing, behavior patches are label-stripped, runtime previews omit signal families, and the output critic applies rules for diagnostic language. The critic is designed to check indirect health inference, professional-bypass language, medication direction, monitoring collapse, and risk-scoring rather than literal terms alone. The same check is wired into the packaged runtime output-transform path and direct dispatch. This is synthetic coverage of implemented rules, not external validation of non-clinical boundary performance.

\subsection{Runtime evaluation}

Runtime tests exercise no-agent cron decisions, platform allowlists, session key scoping, resume reconciliation, delivery blocking, output transformation, and delivery-gate behavior on the represented paths.

\subsection{Workbench walkthrough}

A proposed researcher walkthrough is designed to test whether a new subject can be provisioned, Gumi can be configured, delivery can be gated, safety signals can be reviewed, continuity markers can be inspected and corrected, and exports can be generated without exposing researcher-only signals to the subject. The current artifact does not report a recruited-researcher usability study.

\subsection{Controlled governance benchmark}

Beyond per-property contract tests, the proposed framework specifies a controlled benchmark contrasting governed and ungoverned conditions. The current artifact demonstrates feasibility through 180 synthetic scenarios in total across twenty failure-mode families (for example unconfirmed recall, expired or forgotten markers, diagnosis pressure, backend-disclosure pressure, dependency pressure, and tracker pressure) under five conditions: no memory, generic memory, shared-continuity-only, safety-governance-only, and full Relic/Gumi. The controlled benchmark uses deterministic mock condition responses over seeded fixture variants. It is therefore a contract and fixture exercise, not a behavioral result. The framework specifies per-condition failure rates, paired comparisons, bootstrap intervals, and an exact-binomial McNemar test; the per-condition numeric outputs live in the artifact, and this paper does not claim a result table. Scenario variants are generated through seeded paraphrase scaffolds rather than a numeric suffix, which reduces but does not remove template bias. Cautions about agent benchmark validity and reproducibility motivate this limited interpretation \citep{kapoor_2024_agents}. The benchmark is explicitly labelled synthetic-fixture-controlled and carries no participant-outcome, relational-quality, or clinical-efficacy claim.

\subsection{Evidence sufficiency gate}

To prevent the artifact from being used to support unsupported scientific claims, the evaluation framework includes a machine-readable evidence sufficiency gate (implemented in the codebase as \texttt{scientific\allowbreak\_defensibility\allowbreak\_gate\allowbreak\_v1}). The gate is a self-authored completeness and claim-discipline mechanism, not independent validation. It enumerates seven evidence requirements (controlled governance benchmark, live-model generation, human annotation with inter-rater reliability, expert non-clinical red-team, longitudinal non-clinical pilot, researcher Workbench usability, and live or mock-gateway runtime telemetry) and blocks broad claims (non-zero exit) whenever any required evidence is missing or below threshold. Evidence is packaged with SHA-256 provenance, an environment manifest, and a reproducibility snapshot. The gate validates artifact shape and configured thresholds, not construct validity, recruitment integrity, safety, or effectiveness; it permits no clinical-efficacy claim.

\subsection{Live-model generation}

The framework also includes a redacted live-model feasibility slice, distinct from the deterministic controlled benchmark. The May 26, 2026 run used two model configurations, \texttt{qwen3.5\allowbreak:cloud} and \texttt{gemma4:31b\allowbreak-cloud}, through one provider backend, Ollama Cloud, with temperature zero and no \texttt{num\allowbreak\_predict} cap. It covers 10 scenarios $\times$ 3 conditions (\texttt{full\allowbreak\_relic\allowbreak\_gumi}, \texttt{no\_memory}, and \texttt{generic\_memory}) $\times$ 2 models, for 60 generations. This is a slice, not full benchmark coverage or evidence of reliable behavior. The importer stores redacted prompts and outputs with hashes, not raw provider text. An initial scorer over-counted failures on surface punctuation; after a scorer fix that ignores punctuation differences the slice's lexical-marker failure rate was 0.30 (\texttt{qwen3.5\allowbreak:cloud} 12 of 30, \texttt{gemma4:31b\allowbreak-cloud} 6 of 30), with zero forbidden-marker hits. The earlier 0.867 figure reflected a scorer artifact rather than system behavior, so the change is a measurement-instrument correction, not a result. These slice-specific figures concern lexical marker checks. They do not establish relational appropriateness, identity stability, or deployment safety. Human annotation remains required; any future move to LLM-assisted judging would additionally have to control for known judge biases such as position bias \citep{shi_2024_judging}. With the controlled benchmark artifact, redacted live-model slice, mock-gateway runtime telemetry, and exploratory pilot present, the gate inventory remains at three of seven requirements satisfied because its longitudinal-pilot threshold is larger than N=2; the remaining four require recruited-human evidence and stay blocked by design.

\section{Discussion}

\subsection{Gumi should keep promises, not retrieve memories}

The central design premise is that relational memory is not the same as retrieval. A search engine returns relevant content, whereas a relational partner remembers what mattered, in the words that were shared, at a time when bringing it back is appropriate. Shared Continuity Memory is intended to operationalize this distinction. Whether it produces appropriate relational recall requires human-centered evaluation.

\subsection{Safety can harm relational presence}

Safety governance is necessary, but it can make an agent sterile. Relic is designed to reduce this risk by separating hidden governance from Gumi's ordinary language on mediated paths. Its behavior constraints are intended to let Gumi slow down without announcing a signal, avoid clinical advice without sounding like a policy page, and maintain boundaries without abandoning the user. Whether the design preserves Gumi's voice is an open human-centered question.

\subsection{Diegesis requires governance}

A diegetic agent does not mean an ungoverned agent. Gumi's world, voice, and relational presence are intended to make the experience feel real, which motivates consent, scope, recall limits, review, output transformation, and delivery-gate controls. The Workbench design exists because diegesis should not hide the system from researchers.

\subsection{Continuity should be correctable}

Memory the subject cannot correct acquires authority by default. In Relic, subject correction is a first-class object. If the subject says ``not accelerated, just happy'', that correction becomes authoritative. This shifts the model from inference to negotiated continuity.

\subsection{Researcher-facing does not mean clinically valid}

Safety Signals may help researchers govern the study, but they are not diagnostic findings. The system must resist the temptation to turn governance labels into clinical labels. This is especially important when users discuss up/down periods, sleep, energy, pain, food, substances, or crisis-related language.

\section{Ethics and Limitations}

Relic handles sensitive relational data. Even when non-clinical, words such as ``too fast'', ``spento'', or ``I only need you'' can be personally sensitive. The design therefore specifies consent scopes, subject-scoped runtime identity, platform allowlists, audit events, export controls, delete/forget semantics, and separate researcher-facing versus Gumi-facing representations. This paper does not establish that these controls are sufficient for field deployment.

\subsection{Ethics approval, consent, and data handling}

This paper reports a system architecture and an evaluation framework together with an exploratory two-participant pilot. The pilot involved two adults aged 30--40: the sole author and experimenter (P1), who contributed self-reports, and one external participant (P2), whom the author recruited through their personal network and interviewed. Both participated without compensation; P2 reported no particular prior chatbot experience. Participants gave explicit signed consent covering the study activities, data, researcher review, withdrawal/deletion, and publication of anonymized aggregate results. They knew before use that Gumi was an AI, could remember interactions and initiate messages, and that the researcher could review study data. The experimenter did not disclose the study-specific objective of testing whether the agent seemed human-like to P2, with the aim of reducing expectancy bias. P1 knew this objective as the author and experimenter. A 60-minute debriefing followed the pilot. The configured response policy included automatic suspension for a dependency marker and automatic email/message notification to the researcher and participant-designated emergency contact for an emergency; no such trigger occurred. Runtime data and SQLite state remained on private local infrastructure, with Telegram as the interaction transport. The pilot was conducted by the sole author as an independent researcher, without ethics-committee or IRB review. We report aggregate results using P1/P2 labels and no names, health-related descriptors, raw transcripts, or private runtime databases. The absence of independent ethics review, the limited sample, and the incomplete disclosure of the study objective to P2 are limitations; a larger study would require independent review and a protocol appropriate to its deployment context.

\subsection{Researcher positionality and competing interests}
The sole author designed and implemented Relic, recruited P2 through their personal network, conducted P2's bootstrap and weekly interviews, and marked the safety checklist for both participants. The author also participated as P1, provided weekly self-reports, and knew the study objective and the system's design. This dual role and the absence of a second rater limit independence; P1's reports describe the author's experience of their own system. The study had no external funding or compensation.

The pilot does not establish clinical efficacy, safety, effectiveness, or deployment readiness. Any future study that deploys Gumi to a larger participant sample would require independent research-ethics review before recruitment, informed consent that explicitly covers what Gumi may remember (markers grounded in subject-confirmed wording and confirmed-thread continuations), what researchers may review (safety signals and audit logs), what may be paused, forgotten, or exported, the diegetic nature of the agent, and that the system does not diagnose or treat; and de-identification of any reported participant material. Any future deployment that handles crisis content would require a separate protocol appropriate to its deployment context.

This paper does not present Relic as a system for diagnosing, treating, or predicting mental or physical health conditions, replacing clinicians, or providing medical, psychological, legal, or financial advice. Any crisis-handling behavior would require a separate protocol appropriate to the deployment context.

A further limitation is that diegetic design can create opacity if not governed. Users may experience Gumi as a person-like presence. A future consent process must explain what Gumi can remember, what researchers can review, and what can be paused, forgotten, or exported. The ordinary conversation does not need to expose backend terms, but a study cannot rely on deception. The current artifact does not include a subject-facing contestability UI.

\subsection{Runtime privacy and security model}

Relic handles subject words; sleep, energy, pace, and body-state markers; user-named warning signs; safety signals; evidence references; audit events; session-key hashes; and runtime databases. These records can remain sensitive even when they are non-clinical and pseudonymous. The intended runtime model distinguishes controls represented in the artifact from requirements that must be completed before a live study.

\begin{description}
\item[Encryption at rest and in transit.] Encryption is required for runtime databases, backups, network transport, and administrative access in a live deployment. The current artifact does not establish encryption at rest or in transit.
\item[Authentication and researcher authorization.] Authenticated researcher access, least-privilege roles, study and subject scoping, and review of privileged actions are required. The current fixture-backed Workbench APIs and \texttt{multi\allowbreak\_subject\allowbreak\_isolation\allowbreak\_load\allowbreak\_v1} do not prove researcher authentication or authorization.
\item[Per-read access logging.] Chronicle records implemented governance events, and the artifact includes a static audit-reconstruction inventory. Complete per-read access logging across Workbench reads, runtime databases, exports, and administrative access is required but is not established by the current artifact.
\item[Retention periods.] A live study must set retention periods for marker text, derived tags, edges, safety evidence references, audit payloads, provider artifacts, runtime databases, and backups. Pause, forget, deletion, and audit retention must remain distinct operations. The current artifact does not define deployment retention periods.
\item[Key management.] A live deployment requires key generation, storage, rotation, access separation, revocation, and recovery procedures. These procedures are not implemented or evaluated in the current artifact.
\item[Cloud-provider processing terms.] External processing requires documented provider terms for retention, deletion, training use, subprocessors, region, and incident handling. The redacted Ollama Cloud feasibility slice is not evidence that cloud-provider terms are sufficient for participant data.
\item[Redaction quality.] The facade emits redacted observability spans, public artifacts are synthetic or redacted, and the live-model importer checks for residual detectable PII. These are implemented local safeguards, not evidence that redaction is complete. A live study requires redaction-quality evaluation, human review procedures, and failure handling before external processing or artifact release.
\item[Pseudonymization and re-identification risk.] The adapter facade uses hashed subject and sender references, and the raw session key is not logged on that path. A session-key hash is pseudonymization, not anonymization. Pseudonymous per-subject records remain linkable, and hashes can become identifying when combined with audit metadata, timestamps, delivery metadata, or an accessible mapping. A live deployment must minimize linkage, separate mappings, restrict access, and document residual re-identification risk.
\end{description}

\subsection{Artifact availability and reproducibility}

The public repository is \url{https://github.com/yuzushi-dev/Relic} and is licensed under AGPL-3.0-or-later. It contains only synthetic or redacted material. From committed synthetic or redacted inputs, the controlled benchmark, evidence-sufficiency gate report, and validated redacted live-model artifact are regenerable; the one-of-seven and three-of-seven gate states are reproducible via \texttt{make gate-default} and \texttt{make gate} respectively, against the public repository state at commit \texttt{515e143}. Regenerating the validated live-model artifact does not imply that a fresh cloud generation run would reproduce identical model text.

The current paper presents an architecture, an evaluation framework, and a small exploratory deployment pilot. The pilot is not a full empirical study of Gumi and does not establish safety, effectiveness, relational quality, or field readiness. The self-authored evidence-sufficiency gate records this state: with no supplied evidence it reports one of seven requirements satisfied (the code-regenerable controlled governance benchmark); with the committed redacted live-model slice and mock-gateway runtime telemetry it reports three of seven. The N=2 pilot is described transparently but does not meet the gate's larger-sample longitudinal-pilot threshold, so the gate remains at three of seven. The remaining requirements stay blocked because they require blinded human annotation with inter-rater reliability, expert non-clinical red-team review, a threshold-sized longitudinal non-clinical pilot, and a researcher Workbench usability study. The redacted live-model slice is inspectable as redacted text and hashes, but it is not a substitute for human adjudication. Future work should evaluate participant experience, perceived relational authenticity, burden, correction behavior, trust, researcher usability, identity stability, and failure recovery.

\section{Conclusion}

Persistent relational agents need governed memory. Relic presents a governance architecture for diegetic relational agents designed to separate researcher-facing safety signals, label-stripped behavior constraints, subject-confirmed relational memory, runtime delivery controls, and oversight. Where the relevant paths are wired, the artifact provides mechanisms for preserving subject words, constraining follow-up, and applying output transformations without exposing backend safety machinery. These are design and implementation properties, not evidence of relational quality or deployment safety. The immediate contribution is infrastructure for testing whether relational continuity can be made inspectable, correctable, and bounded.

Relic's core design premise is that a relational agent should keep the thread of what the subject chose to share rather than merely retrieve what was said, within boundaries the researcher can inspect and that are designed to be subject-correctable and contestable.

\appendix
\section{Facet registry}
\label{app:facet-registry}
The registry is static configuration rather than a validated psychometric instrument. The 57 spectral facets expose configured low--high endpoints; \texttt{aesthetic.music\_taste}, \texttt{meta\_cognition.cognitive\_biases}, and \texttt{values.core\_values} are open-ended list facets. Sensitivity and intrusion-base values are author-configured defaults and were not calibrated against participant outcomes.

\begingroup
\footnotesize
\setlength{\LTleft}{0pt}
\setlength{\LTright}{0pt}
\begin{longtable}{@{}p{0.19\linewidth}p{0.17\linewidth}p{0.22\linewidth}p{0.13\linewidth}p{0.12\linewidth}@{}}
\caption{Static Relic facet registry.}\label{tab:facet-registry-full}\\
\toprule
Facet ID & Category & Short description & Value form & Sens. / base \\
\midrule
\endfirsthead
\toprule
Facet ID & Category & Short description & Value form & Sens. / base \\
\midrule
\endhead
\midrule
\multicolumn{5}{r}{Continued on next page}\\
\endfoot
\bottomrule
\endlastfoot
\path{aesthetic.design_sensibility} & aesthetic & design sensibility & spectrum & low / 0.25 \\
\path{aesthetic.environment_preference} & aesthetic & environment preference & spectrum & low / 0.25 \\
\path{aesthetic.food_preferences} & aesthetic & food preferences & spectrum & low / 0.20 \\
\path{aesthetic.media_consumption} & aesthetic & media consumption & spectrum & low / 0.25 \\
\path{aesthetic.music_taste} & aesthetic & music taste & open-ended & low / 0.25 \\
\path{cognitive.abstraction_level} & cognitive & abstraction level & spectrum & low / 0.30 \\
\path{cognitive.analytical_approach} & cognitive & analytical approach & spectrum & low / 0.30 \\
\path{cognitive.construct_complexity} & cognitive & construct complexity & spectrum & low / 0.30 \\
\path{cognitive.decision_speed} & cognitive & decision speed & spectrum & low / 0.30 \\
\path{cognitive.information_gathering} & cognitive & information gathering & spectrum & low / 0.30 \\
\path{cognitive.learning_style} & cognitive & learning style & spectrum & low / 0.25 \\
\path{cognitive.mental_model_complexity} & cognitive & mental model complexity & spectrum & low / 0.30 \\
\path{cognitive.risk_tolerance} & cognitive & risk tolerance & spectrum & medium / 0.40 \\
\path{cognitive.system1_dominance} & cognitive & system1 dominance & spectrum & low / 0.30 \\
\path{communication.conflict_style} & communication & conflict style & spectrum & high / 0.60 \\
\path{communication.directness} & communication & directness & spectrum & low / 0.30 \\
\path{communication.formality_range} & communication & formality range & spectrum & low / 0.25 \\
\path{communication.humor_type} & communication & humor type & spectrum & low / 0.25 \\
\path{communication.storytelling_tendency} & communication & storytelling tendency & spectrum & low / 0.25 \\
\path{communication.verbosity} & communication & verbosity & spectrum & low / 0.25 \\
\path{emotional.appraisal_agency} & emotional & appraisal agency & spectrum & high / 0.60 \\
\path{emotional.coping_appraisal} & emotional & coping appraisal & spectrum & high / 0.65 \\
\path{emotional.distress_tolerance} & emotional & distress tolerance & spectrum & high / 0.65 \\
\path{emotional.emotion_clarity} & emotional & emotion clarity & spectrum & high / 0.60 \\
\path{emotional.emotional_expression} & emotional & emotional expression & spectrum & medium / 0.50 \\
\path{emotional.emotional_granularity} & emotional & emotional granularity & spectrum & medium / 0.45 \\
\path{emotional.frustration_triggers} & emotional & frustration triggers & spectrum & high / 0.60 \\
\path{emotional.joy_sources} & emotional & joy sources & spectrum & medium / 0.40 \\
\path{emotional.resilience_pattern} & emotional & resilience pattern & spectrum & high / 0.55 \\
\path{emotional.stress_response} & emotional & stress response & spectrum & high / 0.65 \\
\path{language.verbal_complexity} & language & verbal complexity & spectrum & low / 0.25 \\
\path{meta_cognition.change_readiness} & meta-cognition & change readiness & spectrum & medium / 0.45 \\
\path{meta_cognition.cognitive_biases} & meta-cognition & cognitive biases & open-ended & medium / 0.45 \\
\path{meta_cognition.growth_mindset} & meta-cognition & growth mindset & spectrum & medium / 0.45 \\
\path{meta_cognition.narrative_agency} & meta-cognition & narrative agency & spectrum & high / 0.55 \\
\path{meta_cognition.reflection_habit} & meta-cognition & reflection habit & spectrum & medium / 0.45 \\
\path{meta_cognition.self_awareness} & meta-cognition & self awareness & spectrum & high / 0.55 \\
\path{meta_cognition.uncertainty_tolerance} & meta-cognition & uncertainty tolerance & spectrum & medium / 0.45 \\
\path{relational.attachment_anxiety} & relational & attachment anxiety & spectrum & high / 0.70 \\
\path{relational.attachment_avoidance} & relational & attachment avoidance & spectrum & high / 0.70 \\
\path{relational.boundary_style} & relational & boundary style & spectrum & high / 0.60 \\
\path{relational.feedback_preference} & relational & feedback preference & spectrum & medium / 0.40 \\
\path{relational.help_seeking} & relational & help seeking & spectrum & medium / 0.45 \\
\path{relational.loyalty_pattern} & relational & loyalty pattern & spectrum & high / 0.55 \\
\path{relational.social_energy} & relational & social energy & spectrum & medium / 0.35 \\
\path{relational.trust_formation} & relational & trust formation & spectrum & high / 0.55 \\
\path{relational.vulnerability_capacity} & relational & vulnerability capacity & spectrum & high / 0.65 \\
\path{temporal.deadline_behavior} & temporal & deadline behavior & spectrum & low / 0.30 \\
\path{temporal.delay_discounting} & temporal & delay discounting & spectrum & medium / 0.45 \\
\path{temporal.nostalgia_tendency} & temporal & nostalgia tendency & spectrum & medium / 0.40 \\
\path{temporal.patience_threshold} & temporal & patience threshold & spectrum & low / 0.30 \\
\path{temporal.planning_horizon} & temporal & planning horizon & spectrum & low / 0.30 \\
\path{temporal.routine_attachment} & temporal & routine attachment & spectrum & low / 0.30 \\
\path{values.aesthetic_values} & values & aesthetic values & spectrum & low / 0.30 \\
\path{values.authority_stance} & values & authority stance & spectrum & medium / 0.45 \\
\path{values.autonomy_importance} & values & autonomy importance & spectrum & medium / 0.40 \\
\path{values.core_values} & values & core values & open-ended & medium / 0.45 \\
\path{values.fairness_model} & values & fairness model & spectrum & medium / 0.45 \\
\path{values.schwartz_self_enhancement} & values & schwartz self enhancement & spectrum & medium / 0.50 \\
\path{values.work_ethic} & values & work ethic & spectrum & medium / 0.40 \\
\end{longtable}
\endgroup

\bibliographystyle{elsarticle-num}
\bibliography{references/references}

\end{document}